# Unified Growth Theory Contradicted by the Mathematical Analysis of the Historical Growth of Human Population

Ron W Nielsen[1]

Griffith University, Environmental Futures Research Institute, Gold Coast Campus, Australia

**Abstract.** Data describing the historical growth of human population global and regional (Western Europe, Eastern Europe, Asia, former USSR, Africa and Latin America) are analysed. Results are in harmony with the earlier analysis of the historical growth of the world population in the past 12,000 years and with a similar but limited study carried out over 50 years ago. This analysis is also in harmony with the study of the historical economic growth. Within the range of analysable data, there was no Malthusian stagnation. Takeoffs from stagnation to growth, postulated by the Unified Growth Theory never happened. There were no escapes from the Malthusian trap because there was no trap. This analysis and the earlier studies of the Gross Domestic Product lead to the conclusion that there were also no takeoffs in the income per capita distributions, claimed by the Unified Growth Theory. Consequently, the claimed in this theory differential timing in takeoffs never happened. Unified Growth Theory is contradicted yet again by the mathematical analysis of the same data, which were used, but never analysed, during the formulation of this theory. However, this study, as well as the earlier publications on the related topics, shows also that some fundamental postulates used in the economic and demographic research are repeatedly contradicted by the mathematical analysis of data.

## Introduction

Historical economic growth can be studied using the Gross Domestic Product (GDP). However, to understand the time dependence of the income per capita (GDP/cap) it is necessary to understand not only the economic growth, expressed in terms of the GDP, but also the growth of human population. We have already analysed the GDP data (Nielsen, 2016a). Now, we shall analyse the growth of human population using the same source of data (Maddison, 2010). The aim of all these studies is to investigate the validity of the Unified Growth Theory (Galor, 2005a, 2011).

Our earlier analysis (Nielsen, 2016a) demonstrated that the historical economic growth, regional and global, was following hyperbolic distributions. Analysis published over 50 years ago (von Foerster, Mora & Amiot, 1960) demonstrated that the growth of the world population was also hyperbolic during the AD era. The follow-up analysis (Nielsen, 2016c) demonstrated that the growth of the world population was hyperbolic not only during the AD era but also during the BC era, for the total of around 12,000 years. This particular analysis identified two demographic transitions in the past growth of the population: between 500 BC and AD 500 and between AD 1200 and 1400. However, these transitions were of a different kind than transitions used routinely in demographic research. They were not transitions from stagnation to growth but from growth to growth, or more precisely, from hyperbolic growth to hyperbolic growth. The first transition was from a fast hyperbolic growth (as defined by the parameter $k$, explained below) during the BC era to a significantly slower hyperbolic growth during the AD era. During this transition, the size of human population reached a maximum around AD 1 and after reaching a minimum between AD 400 and 500 it resumed it slower hyperbolic growth during the AD era. However, the starting size of the population in AD 500 was significantly larger than in 10,000 BC and the slower hyperbolic growth increased rapidly to reach a large size of the population in only about 2000 years. During this first demographic transition, the growth rate decreased from 0.252% in 500 BC to 0.066% in AD 500. The second transition was hardly noticeable but it resulted in a change from a slow hyperbolic trajectory to slightly faster hyperbolic trajectory. During this transition, after a short delay in the growth of the population, the growth rate increased only marginally from 0.123% in AD 1200 to 0.157% in AD 1400. Currently the growth of the world population experiences a third demographic transition to a yet unknown trajectory.

## Unified Growth Theory

The latest and the most elaborate theory describing economic growth is the Unified Growth Theory (Galor, 2005a, 2011). It is not a theory, which is widely accepted by economist and used in their research. In fact, the opposite seems to be true. However, we are using this theory *as an example* for two reasons. *First*, it is a theory, which is firmly supported by traditional assumptions about the historical economic growth and about the historical growth of human population, assumptions, which are based on strongly questionable conjectures. Our primary aim here, as well as in our earlier publications (Nielsen, 2014, 2015a, 2016a, 2016b, 2016c, 2016d, 2016e), is not just to test the validity of this theory or the validity of a similar Demographic Transition Theory (see Nielsen, 2016e and references therein) but to test the validity of the *fundamental postulates* used in economic and demographic research. *Second*, Unified Growth Theory appears to be the only theory where Maddison's data (Maddison, 2001) were systematically used.

In the last years of his life, Magnusson, the world-renown economist, published excellent data describing not only the economic growth as expressed by the Gross Domestic Product (GDP) but also the growth of human population, global, regional and national (Magnuson, 2001, 2010). These data are a treasure trove, which can be used in the economic and demographic research. In particular, they can be used to test the fundamental postulates supporting these two fields of research. Galor used the earlier compilation of these data (Magnuson, 2001) but any of these compilations can be used to test the fundamental postulates supporting economic and demographic research, and in particular to test the validity of the Unified Growth Theory.

Unfortunately, Galor did not use Maddison's data in the way they should be used in scientific research. He did not analyse data. He did not use these data to *test* the fundamental assumptions but to *confirm* them. Such a use of data might be tolerable in disciplines where doctrines are accepted by faith but not in science. To this end, data were systematically manipulated by repeatedly quoting, for instance, some well-chosen and isolated figures. His theory and its fundamental postulates are also based on the habitually distorted and self-misleading presentations of data (Ashraf, 2009; Galor, 2005a, 2005b, 2007, 2008a, 2008b, 2008c, 2010, 2011, 2012a, 2012b, 2012c; Galor and Moav, 2002; Snowdon & Galor, 2008). This approach to research was used to promote such scientifically-unsupported concepts as the concept of the three regimes of growth (Malthusian regime of stagnation, post-Malthusian regime and sustained-growth regime), the concept of sudden takeoffs from stagnation to growth, the concept of differential takeoffs and the concept of the great divergence. An example of such diagrams is shown in Figure 1. (All diagrams are presented in the Appendix.)

Hyperbolic distributions do not have to be distorted to be confusing. They are already sufficiently confusing and it is easy to make mistakes with their interpretations. Hyperbolic distributions have to be carefully and methodically analysed and fortunately their analysis becomes trivial when using the reciprocal values of data (Nielsen, 2014). Displays, such as presented in Figure 1, which is based on a figure presented by Galor (2005a, p. 181), are self-misleading and they inevitably lead to incorrect conclusions.

The correct and accurate display of Maddison's data (Maddison, 2001), precisely the same data as used but never scientifically analysed during the formulation of the Unified Growth Theory (Galor, 2005a, 2011), is presented in Figure 2. Analysis of these data reveals that they follow monotonically-increasing distributions, which are impossible to divide into distinctly-different components governed by distinctly-different mechanisms of growth (Nielsen, 2015a).

Whether expressed by using the GDP or GDP/cap, the economic growth was slow over a long time and fast over a short time but it was monotonically increasing all the time. What appears as stagnation was a part of the monotonically-increasing distribution, and what appears as a sudden takeoff was the natural continuation of the same monotonically-increasing distribution.

Attempts to determine the time of the perceived transition from the perceived slow to a fast growth are bound to be unsuccessful because there was no transition (Nielsen, 2014, 2015a). The growth of the GDP is described by hyperbolic distributions (Nielsen, 2016a) and the growth of the GDP/cap by the linearly-modulated hyperbolic distributions (Nielsen, 2015a).

We have already demonstrated that the Unified Growth Theory is contradicted by the GDP data for Western Europe, Eastern Europe, countries of former USSR, Asia, Africa and Latin America (Nielsen, 2016b, 2016d). We have also demonstrated that the Unified Growth Theory is contradicted by the data describing the growth of the world income per capita (Nielsen, 2015a). Our next step now it to investigate the properties of regional growth of income per capita (GDP/cap). To this end we have to analyse first regional population data (Maddison, 2010) but we shall also include the analysis of the world population data.

One of the fundamental postulates of the Unified Growth Theory is the postulate of the existence of three regimes of growth governed by three distinctly different mechanisms: (1) the Malthusian regime of stagnation, (2) the post-Malthusian regime, and (3) the sustained-growth regime. This postulate applies not only to the growth of the GDP but also to the growth of human population because Galor discusses the growth of income per capita, (GDP/cap), which is made of two components: the growth of the GDP and the growth of the population.

According to Galor (2005a, 2008a, 2011, 2012a), Malthusian regime of stagnation was between 100,000 BC and AD 1750 for developed regions and between 100,000 BC and AD 1900 for less-developed regions. The claimed starting time appears to be based entirely on conjecture because Maddison's data are terminated at AD 1 and even they contain significant gaps below AD 1500. The post-Malthusian regime was allegedly between AD 1750 and 1850 for developed regions and from 1900 for less-developed regions. The sustained-growth regime was supposed to have commenced around 1850 for developed regions.

Unified Growth Theory (Galor, 2005a, 2008a, 2011, 2012a) can be tested in many ways but the easiest way to test it is to look for the dramatic takeoffs from stagnation to growth. These takeoffs are described as a "remarkable" or "stunning" escape from the Malthusian trap (Galor, 2005a, pp. 177, 220). It is a signature, which cannot be missed.

This change in the pattern of growth is described as "the sudden take-off from stagnation to growth" (Galor, 2005a, pp. 177, 220, 277) or as a "sudden spurt" (Galor, 2005a, 177, 220). According to Galor, for developed regions, the end of the Malthusian regime of stagnation coincides with the Industrial Revolution. "The take-off of developed regions from the Malthusian Regime was associated with the Industrial Revolution" (Galor, 2005a, p. 185). Indeed, the Industrial Revolution is considered to have been "the prime engine of economic growth" (Galor, 2005a, p. 212).

This signature is characterised by three features: (1) it should be a prominent change in the pattern of growth, (2) it should be a transition from stagnation to growth and (3) it should occur at the time predicted by the theory. For developed regions, the postulated takeoffs should occur around AD 1750, or around the time of the Industrial Revolution, 1760-1840 (Floud & McCloskey, 1994). For less-developed regions, they should occur around 1900. The added advantage of using this simple test is that there are no significant gaps in the data around the time of the postulated takeoffs and consequently the stagnation and the expected prominent transitions from stagnation to growth should be easily identifiable.

A transition from growth to growth is not a signature of the postulated takeoff from stagnation to growth. Thus, for instance, a transition from hyperbolic growth to another hyperbolic growth or to some other steadily-increasing trajectory is not a signature of the sudden takeoff from stagnation to growth. Likewise, a transition at a distinctly different time is not a confirmation of the theoretical expectations.

The takeoffs claimed by Galor are in the income per capita (GDP/cap), which means that there should be takeoffs from stagnation to growth in at least one of these components (in the GDP or in the population or in both of them) at a specific time (Galor, 2008a, 2012a). We have already demonstrated that there were no takeoffs in the growth of the GDP (Nielsen, 2016d). Consequently, to confirm the Unified Growth Theory we would have to show not only that there were takeoffs from stagnation to growth in the growth of the population but also that these takeoffs occurred at the specific time claimed by Galor (2008a, 2012a), around AD 1750 for developed regions (Western Europe, Eastern Europe and the former USSR) and at around AD 1900 for less developed regions (Asia, Africa and Latin America). We shall now demonstrate that there were no such takeoffs. Thus we shall demonstrate implicitly that there were no takeoffs in the income per capita, which means that Galor's postulate of the differential timing in takeoffs is also contradicted by data, because we cannot have differential timing in takeoffs without takeoffs.

## Essentials of the mathematical analysis

Hyperbolic distribution describing growth is represented by a reciprocal of a linear function:

$$S(t) = \frac{1}{a - kt}, \quad (1)$$

where $S(t)$ is the size of the growing entity, in our case the population, while $a$ and $k$ are positive constants.

As pointed out earlier (Nielsen, 2014), hyperbolic distributions are confusing because they create an illusion of being made of two components, slow and fast, with perhaps even a third component in the middle. It is easy to make a mistake with their interpretations. Fortunately, these distributions are easy to analyse by using the reciprocal values of data, $1/S(t)$:

$$\frac{1}{S(t)} = a - kt. \quad (2)$$

In this representation, data follow a decreasing straight line, which obviously cannot be divided into three distinctly different components.

Reciprocal values help in an easy and generally unique identification of hyperbolic growth. Apart from serving as an alternative way to analyse data, reciprocal values allow also for the investigation of even small deviations from hyperbolic distributions because deviations from a straight line can be easily noticed.

The illusion of different components also disappears when using semilogarithmic scales of reference. Both types of displays help in an easy identification of disagreements between data and fitted curves for small values of data and we shall use both of these displays.

## Growth of the world population

Results of mathematical analysis of the world economic growth are presented in Figures 3 and 4. Reciprocal values of historical data identify uniquely hyperbolic distribution between AD 1000 and around 1950 because the reciprocal data follow a decreasing straight line. From around 1950, the growth of the world population started to be diverted to a slower trajectory but first it was slightly boosted. The boosting was small (it is hardly noticeable in the displayed diagrams) and it did not last long.

Hyperbolic fit to the world population data (Maddison, 2010) is shown in Figure 4. The fit is remarkably good. The point at AD 1 is 75% away from the fitted curve. This discrepancy is in perfect agreement with the analysis of the growth of the world population over the past 12,000 years (Nielsen, 2016c), which demonstrated a maximum around that year.

Parameters describing hyperbolic trajectory fitting the data between AD 1000 and 1950 are: $a = 7.739 \times 10^0$ and $k = 3.765 \times 10^{-3}$. Its singularity is at $t = 2056$. However, from around 1950, the growth of the world population started to be diverted to a slower trajectory bypassing the singularity by a safe margin of 106 years. This diversion was first manifested in a minor and short-lasting boosting of the growth of the world population.

The data are in disagreement with the Unified Growth Theory (2005a, 2011). Industrial Revolution had no impact on the growth trajectory. There were also no takeoffs from stagnation to growth around AD 1750 for developed regions and around AD 1900 for less-developed regions because there was no stagnation and because hyperbolic growth continued undisturbed. If there were such takeoffs in the respective regions, we would have expected to see clear distortions of the growth trajectory, but the trajectory was remarkably stable during these alleged but non-existent takeoffs. Unified Growth Theory is yet again demonstrably contradicted by data.

With the absence of the takeoffs in the growth of the population and with the earlier demonstrated absence of the takeoffs in the growth of the GDP (Nielsen, 2016d), this analysis shows that there were no takeoffs in the income per capita (GDP/cap) distribution. Unified Growth Theory (Galor, 2005a, 2011) is contradicted by data, which were used but never analysed during the formulation of this theory.

## Western Europe

Growth of the population in Western Europe is shown in Figures 5 and 6. Western Europe is represented by the total of 30 countries: Austria, Belgium, Denmark, Finland, France, Germany, Italy, The Netherlands, Norway, Sweden, Switzerland, United Kingdom, Greece, Portugal, Spain and by 14 small, but unspecified countries. Ireland is missing in this list because it was included only from 1921.

The straight line fitting the reciprocal values of data, shown in Figure 5, identifies uniquely hyperbolic distribution between AD 1000 and around 1915. Parameters describing the hyperbolic growth in Western Europe are: $a = 7.542 \times 10^1$ and $k = 3.749 \times 10^{-2}$. The point of singularity is at $t = 2012$. From around 1915, the growth of the population in Western Europe started to be diverted to a slower, but still fast-increasing, trajectory bypassing the singularity by safe margin of 97 years. The size of the population in AD 1 is 89% higher than the fitted hyperbolic distribution. This discrepancy is probably reflecting a maximum in the growth of the world population around that year (Nielsen, 2016c).

Figures 5 and 6 show that hyperbolic growth between AD 1000 and 1915 remained undisturbed. Industrial Revolution had absolutely no impact on changing the hyperbolic growth trajectory in the region where the effects of this revolution should be most prominent. There was no takeoff from stagnation to growth at the postulated time (Galor, 2008a, 2012a) because there was no stagnation but a hyperbolic growth. There was even no transition to a faster hyperbolic trajectory.

With the absence of the takeoff in the growth of the population in Western Europe and with the earlier demonstrated absence of the takeoff in the growth of the GDP (Nielsen, 2016d), this analysis shows that there was no takeoff in the income per capita (GDP/cap) distribution. Unified Growth Theory (Galor, 2005a, 2011) is contradicted by data, which were used but never analysed during the formulation of this theory.

## Eastern Europe

Results of analysis of the growth of population in Eastern Europe are summarized in Figures 7 and 8. Reciprocal values of data shown in Figure 7 identify uniquely hyperbolic distribution between AD 1000 and around 1935. From that year, the growth of population started to be diverted to a slower trajectory.

Hyperbolic parameters are: $a = 3.055 \times 10^2$ and $k = 1.525 \times 10^{-1}$. The point of singularity is at $t = 2003$. Figures 7 and 8 demonstrate that the Industrial Revolution had no impact on the trajectory of the growth of the population in Eastern Europe and that there was no takeoff from stagnation to growth at a postulated time (Galor, 2008a, 2012a) because there was no stagnation but hyperbolic growth. There was even no takeoff to a faster hyperbolic growth. The size of the population at AD 1 was 45% higher than the calculated curve reflecting probably the maximum in the growth of the world population around that year (Nielsen, 2016c).

With the absence of the takeoff in the growth of the population in Eastern Europe and with the earlier demonstrated absence of the takeoff in the growth of the GDP (Nielsen, 2016d), this analysis shows that there was no takeoff in the income per capita (GDP/cap) distribution. Unified Growth Theory (Galor, 2005a, 2011) is contradicted by data, which were used but never analysed during the formulation of this theory.

## Former USSR

The analysis of data for the countries of the former USSR is presented in Figures 9 and 10. Reciprocal values shown in Figure 9 identify uniquely hyperbolic distribution between AD 1 and around 1920. The hyperbolic fit to the data is between AD 1 and 1870. Parameters fitting the data are: $a = 2.618 \times 10^2$ and $k = 1.333 \times 10^{-1}$. The singularity is at $t = 1965$ From around 1920, the growth of the population in the former USSR started to be diverted to a slower trajectory, bypassing the singularity by around 45 years.

Figures 9 and 10 show that the Industrial Revolution had no impact on shaping the growth of human population in the countries of the former USSR. There was also no takeoff from stagnation to growth around the postulated time Galor (2008a, 2012a) or around any other time because the growth was not stagnant but hyperbolic. There was even no transition to a faster hyperbolic trajectory but there was a transition to a slower, non-hyperbolic growth around 1920.

With the absence of the takeoff in the growth of the population in the countries of the former USSR and with the earlier demonstrated absence of the takeoff in the growth of the GDP (Nielsen, 2016d), this analysis shows that there was no takeoff in the income per capita (GDP/cap) distribution. Unified Growth Theory (Galor, 2005a, 2011) is contradicted by data, which were used but never analysed during the formulation of this theory.

## Asia

Analysis of the growth of human population in Asia (including Japan) is summarised in Figures 11 and 12. Reciprocal values presented in Figure 11 identify uniquely hyperbolic distribution between AD 1000 and around 1920. The parameters describing this distribution are: $a = 1.068 \times 10^1$ and $k = 4.999 \times 10^{-3}$. The point of singularity is at $t = 2135$.

Asia is made primarily of less-developed countries (BBC, 2014; Pereira, 2011) and consequently, according to Galor (2008a, 2012a), the growth of human population in Asia should have been characterised by stagnation until around 1900, the year marking the alleged stunning escape from Malthusian trap, the escape manifested by the postulated dramatic takeoff. (The population of Japan before AD 1900 was on average less than 4% of the total population of Asia.) The data and their analysis show that there was no stagnation, at least from AD 1000 and no expected takeoff.

The data reveal a steadily increasing hyperbolic growth until around 1920. From around that year the growth of human population was diverted to a faster trajectory. This boosting can be seen clearly in Figures 11 and 12 and it occurred close to the time of the postulated takeoff from stagnation to growth. However, it was not a transition from stagnation to growth but from hyperbolic growth to a slightly faster trajectory of a different kind. It is, therefore, not the takeoff postulated by Galor. Furthermore, it was only a temporary boosting, which is now returning to the original hyperbolic trajectory and, as indicated by the reciprocal values of data, this new growth is likely to be slower than the original trajectory.

With the absence of the postulated takeoff in the growth of the population in Asia and with the earlier demonstrated absence of the takeoff in the growth of the GDP (Nielsen, 2016d), this analysis shows that there was no takeoff in the income per capita (GDP/cap) distribution. Unified Growth Theory (Galor, 2005a, 2011) is contradicted by data, which were used but never analysed during the formulation of this theory.

## Africa

Results of analysis of the growth of human population in the 57 African countries are presented in Figures 13 and 14. Reciprocal values identify uniquely two hyperbolic trajectories: AD 1-1840 and AD 1840-1980. At first it was a slow hyperbolic growth characterised by parameters $a = 5.794 \times 10^1$ and $k = 2.473 \times 10^{-2}$ and by the singularity at $t = 2343$. Then, around 1840, this slow hyperbolic growth was replaced by a significantly faster hyperbolic growth characterised by parameters $a = 1.571 \times 10^2$ and $k = 7.834 \times 10^{-2}$ and by the singularity at $t = 2006$. Defined by the parameter $k$, this new growth was 3.2 times faster than the earlier hyperbolic growth. From around 1980, this fast hyperbolic growth was diverted to a slower, non-hyperbolic trajectory, bypassing singularity by 26 years.

Africa is also made of less-developed countries (BBC, 2014; Pereira, 2011) so according to Galor (2008a, 2012a) it should have experienced stagnation until around 1900 followed by a clear takeoff around that year. These expectations are contradicted by data because (1) the growth of population was not stagnant but hyperbolic until around 1980 and (2) because there was no takeoff from stagnation to growth around 1900 or around any other time. In fact around that time hyperbolic growth continued unaffected in contradiction of the wished-for interpretations.

The acceleration in the growth of human population in Africa occurred around 1840, but it was not a transition from stagnation to growth but from growth to growth. Even more precisely, it was a transition from the hyperbolic growth to another hyperbolic growth.

This acceleration can be probably explained by noticing that it appears to coincide with the intensified colonisation of Africa (Duignan & Gunn, 1973; McKay, Hill, Buckler, Ebrey, Beck, Crowston, & Wiesner-Hanks, 2012; Pakenham, 1992). The fast growth of the population after 1840 was not reflecting the rapidly improving living conditions of African population brought about by the beneficial changes caused by the Industrial Revolution but by the rapidly increasing wealth of new settlers and their countries of origin at the expense of the deplorable living conditions of the native populations because as shown elsewhere (Nielsen, 2013) growth rate of human population is directly proportional to the level of deprivation.

With the absence of the takeoff in the growth of the population in Africa and with the earlier demonstrated absence of the takeoff in the growth of the GDP (Nielsen, 2016d), this analysis shows that there was no takeoff in the income per capita (GDP/cap) distribution. Unified Growth Theory (Galor, 2005a, 2011) is contradicted by data, which were used but never analysed during the formulation of this theory.

## Latin America

Results of analysis of population growth in Latin America are presented in Figures 15 and 16. Data for Latin America are difficult to analyse because there was a significant decline in the growth of the population between AD 1500 and 1600 but they also appear to follow two distinctly different hyperbolic trajectories, which can be easily identified using the reciprocal values of data (see Figure 15). However, the identification of the first trajectory is not as clear as for Africa. The identification of the second hyperbolic trajectory is more convincing. Tentative conclusion is that the growth of population in Latin America was following a slow hyperbolic distribution between AD 1 and 1500 and a fast hyperbolic distribution between AD 1600 and around 1900.

The tentatively assigned slow hyperbolic growth between AD 1 and 1500 is characterised by parameters $a = 1.765 \times 10^2$ and $k = 8.242 \times 10^{-2}$. Its singularity is at $t = 2142$. The better determined fast hyperbolic growth between AD 1600 and 1900 is characterised by parameters $a = 6.561 \times 10^2$ and $k = 3.371 \times 10^{-1}$. Its singularity is at $t = 1947$. Defined by the parameter $k$, this growth was 4.1 times faster than the earlier hyperbolic growth. From around 1900, this fast hyperbolic growth started to be diverted to a slower trajectory bypassing the singularity by 47 years. The transition from the earlier apparent hyperbolic growth to a new and rapid hyperbolic growth, which occurred between around AD 1500 and 1600 appears to coincide with the commencement of the Spanish conquest (Bethell, 1984).

Latin America is also made of less-developed countries (BBC, 2014; Pereira, 2011) so again, according to Galor (2008a, 2012a), the growth of human population in this regions should have been stagnant until around 1900 and fast-increasing from around that year. This pattern of growth is contradicted by data. The data show a diametrically different pattern: (1) there is no convincing evidence of the existence of stagnation over the entire range of time between AD 1 and 1900 but there is a sufficiently convincing evidence of the hyperbolic growth particularly between AD 1600 and 1900; (2) there was no takeoff from stagnation to growth at any time; and (3) at the time of the postulated takeoff in 1900 the growth of the population started to be diverted to a slower trajectory. The wished-for takeoff is replaced by a slower growth. However, even if we had a takeoff around that time it would have been a takeoff of a different kind, not a takeoff from stagnation to growth as required by the Unified Growth Theory (Galor, 2005a, 2011) but a takeoff from growth to growth.

With the absence of the takeoff in the growth of the population in Latin America and with the earlier demonstrated absence of the takeoff in the growth of the GDP (Nielsen, 2016d), this analysis shows that there was no takeoff in the income per capita (GDP/cap) distribution. Unified Growth Theory (Galor, 2005a, 2011) is contradicted by data, which were used but never analysed during the formulation of this theory.

## Summary and conclusions

Results of mathematical analysis of the historical growth of human population are summarised in Table 1. The listed parameters *a* and *k* are for the fitted hyperbolic distributions.

**Table 1.** *Summary of the mathematical analysis or the historical growth of population*

| Region/Countries | *a* | *k* | Hyperbolic Range | Singularity | Proximity | Take off |
|---|---|---|---|---|---|---|
| World | $7.739 \times 10^0$ | $3.765 \times 10^{-3}$ | 1000 – 1950 | 2056 | 106 | X |
| Western Europe | $7.542 \times 10^1$ | $3.749 \times 10^{-2}$ | 1000 – 1915 | 2012 | 97 | X |
| Eastern Europe | $3.055 \times 10^2$ | $1.525 \times 10^{-1}$ | 1000 – 1935 | 2003 | 68 | X |
| Former USSR | $2.618 \times 10^2$ | $1.333 \times 10^{-1}$ | 1 – 1920 | 1965 | 45 | X |
| Asia | $1.068 \times 10^1$ | $4.999 \times 10^{-3}$ | 1000 – 1920 | 2135 | 215 | X |
| Africa | $5.794 \times 10^1$ | $2.473 \times 10^{-2}$ | 1 – 1840 | 2343 | | X |
| | $1.571 \times 10^2$ | $7.834 \times 10^{-2}$ | 1840 – 1980 | 2006 | 26 | |
| Latin America | $1.765 \times 10^2$ | $8.242 \times 10^{-2}$ | 1 – 1500 | 2142 | | X |
| | $6.561 \times 10^2$ | $3.371 \times 10^{-1}$ | 1600 – 1900 | 1947 | 47 | |

**Notes:** *a* and *k* – Hyperbolic growth parameters [see eqn (1)]. *Hyperbolic Range* - The empirically-confirmed range of time when the growth of population can be described using hyperbolic distributions. *Singularity* - The time of the escape to infinity for a given hyperbolic distribution. *Proximity* - Proximity (in years) of singularity at the time when the economic growth departed from the hyperbolic growth to a new trajectory. *X* - No takeoff. The takeoff from stagnation to growth claimed by the Unified Growth Theory (Galor, 2005a, 2008a, 2011, 2012a) did not happen.

This analysis demonstrates that the natural tendency for the historical growth of human population was to increase hyperbolically. In general, there is a remarkably good agreement between the data and the calculated hyperbolic distributions.

Unlike the more familiar exponential distributions, which are easier to understand because they show more readily a gradually increasing growth, hyperbolic distributions appear to be made of two or maybe even three components: a slow component, a fast component and perhaps even a transition component located between the apparent slow and fast components. The illusion is so strong that even the most experienced researchers can be deceived particularly if they have no access to good sets of data, which was in the past. Now, however, excellent data are available (Maddison, 2001, 2010) and we can use them to check the earlier interpretations of economic growth and of the growth of human population.

The postulate of the existence of the epoch of Malthusian stagnation is suggested by a slow growth over a long time but this slow growth is just a part of the hyperbolic growth, which is convincingly identified using reciprocal values. Hyperbolic distributions create also the illusion of a sudden takeoff but this feature is also a part of hyperbolic growth.

Hyperbolic growth is slow over a long time and fast over a short time but the slow and fast growth are the integral features of the same monotonically increasing distribution, which is easier to understand by using the reciprocal values of the growing entity (Nielsen, 2014). In such displays, the illusion of distinctly different components disappears because hyperbolic growth is then represented by a decreasing straight line, which is easy to understand. It then becomes obvious that hyperbolic distribution cannot be divided into distinctly different sections governed by different mechanism because it makes no sense to divide a straight line into arbitrarily chosen sections and claim different mechanism for such arbitrarily-selected section. It is also then clear that it is impossible to determine the transition from a slow to a fast growth. Which point on a straight line should we select to identify such a transition? The transition does not happen at any specific time but gradually over the whole range of time.

Our analysis shows that the Industrial Revolution had no impact on the growth of human population. It also shows that the postulated takeoffs (Galor, 2005, 2008a, 2011, 2012a) never happened. We have shown earlier (Nielsen, 2016d) that there were no takeoffs in the growth of the GDP, global or regional. The demonstrated now absence of takeoffs in the growth of population shows that the claimed by Galor takeoffs in the income per capita (GDP/cap) did not exist.

Galor describes the imaginary and non-existing features, which have nothing to do with the economic growth or with the growth of human population, features which were conjured from such inaccurate displays as shown in Figure 1, interpretations based on impressions, which were never checked by the scientific analysis of data. It is a world of fiction. All his explanations of the mechanism of economic growth based on these and other imaginary features are not only irrelevant but also misleading.

Galor's Unified Growth Theory is fundamentally incorrect and is repeatedly contradicted by data (Nielsen, 2014, 2015a, 2016a, 2016b, 2016d), ironically by the same data, which were used but never analysed during the formulation of this theory. The evidence contradicting the fundamental postulates of the Unfired Growth Theory is overwhelming and further evidence will be presented in the forthcoming publications. This evidence questions not only the fundamental postulates of the Unified Growth Theory but also many similar postulates used traditionally in economic and demographic research, postulates which were based largely on impressions and conjectures but postulates, which are now repeatedly contradicted by the analysis of Maddison's data (Maddison, 2001, 2010).

In science, just one contradicting evidence in data is sufficient to show that incorrect postulates should be rejected and that theories based on such postulates should be revised. In its present form, the Unified Growth Theory is scientifically unacceptable and so are also many traditional interpretations of the historical economic growth and of the growth of human population.

The data and their analysis suggest new lines of research. They suggest that our attention should not be directed towards explaining the mechanism of stagnation and of the sudden takeoffs from stagnation to growth because these features are contradicted by data. What needs to be explained is why the historical economic growth and the growth of human population were hyperbolic and why relatively recently they were diverted to slower trajectories.

The correct understanding of the fundamental concepts of economic growth and of the growth of human population is important not only academically but also for practical reasons because the correct understanding of the past growth may help us to understand the current growth and how it should be controlled. For instance, the study of the past growth leads to the conclusion that the spontaneous economic growth and the growth of population were hyperbolic. It is, therefore, possible that the spontaneous future growth might again become hyperbolic, which would be most undesirable because such a growth contains singularity.

The current economic growth and the growth of population is no longer hyperbolic but it still follows closely the historically-determined hyperbolic trajectories and thus if left alone it might easily converge into spontaneously-preferable hyperbolic distributions. If we could explain the mechanism of the spontaneous hyperbolic growth we might be better equipped to control the future growth.

The concept of the ages-long stagnation in the economic growth and in the growth of human population, generally accepted in the economic and demographic research, is potentially dangerous because it diverts our attention from the correct understanding of the dynamics of economic growth and of the growth of population and thus makes us ill-equipped for finding suitable solutions for controlling the future economic growth and of the growth of population.

In this respect, Unified Growth Theory is unacceptable not only scientifically but also for practical reasons. It claims erroneously that after the ages-long stagnation in the economic growth we have now entered a sustained-growth regime. This concept suggests a prosperous and secure future. However, mathematical analysis of data indicates that the economic growth in the past was stable and sustainable (Nielsen, 2016a) but now its future is no longer secure (Nielsen, 2015b). The sense of security created by the scientifically-unacceptable Unified Growth Theory is replaced by the urgent need to control and regulate economic growth. In one of our future publications we shall also explain why the concept of the great divergence promoted by the Unified Growth Theory is also potentially dangerous.

# Appendix

**The typical distorted presentation of data used repeatedly in the Unified Growth Theory (Galor, 2005a, 2011) and in other related publications**

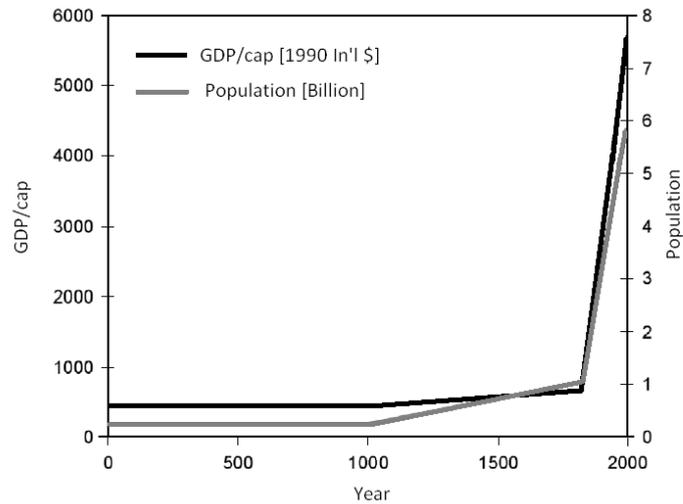

**Figure 1**. *Example of the ubiquitous, grossly-distorted and self-misleading diagrams used to create the Unified Growth Theory (Galor, 2005a, 2011). Madison's data (Maddison, 2001) were used during the formulation of this theory but they were never analysed. Such state-of-the-art was used to construct a system of scientifically-unsupported concepts, interpretations and explanations.*

**The accurate presentation of the same data (Maddison, 2001) and results of their mathematical analysis**

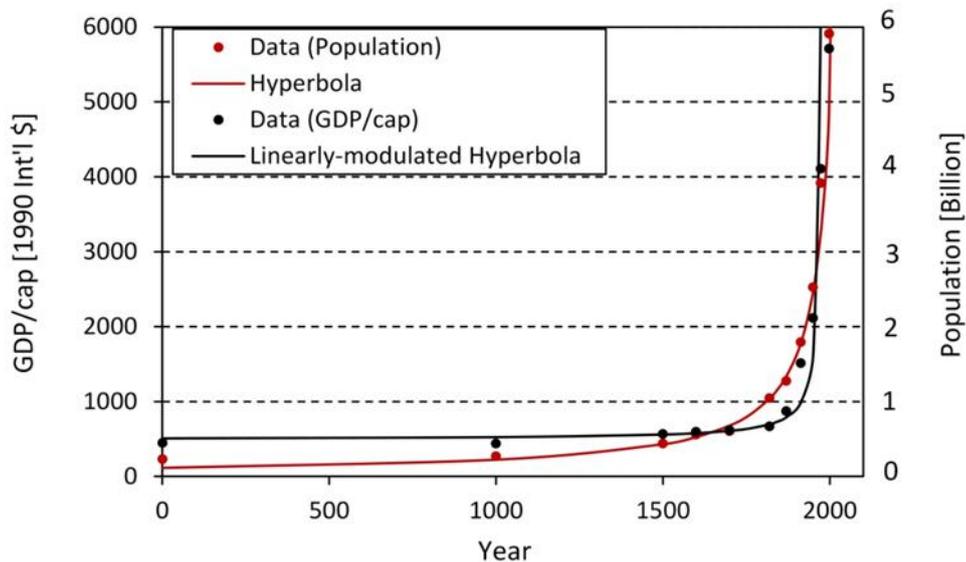

**Figure 2**. *The same data (Maddison, 2001) as used in Figure 1 displayed accurately and analysed. They follow monotonically-increasing distributions, which cannot be divided into distinctively-different components (Nielsen, 2014, 2015a).*

**World Population**

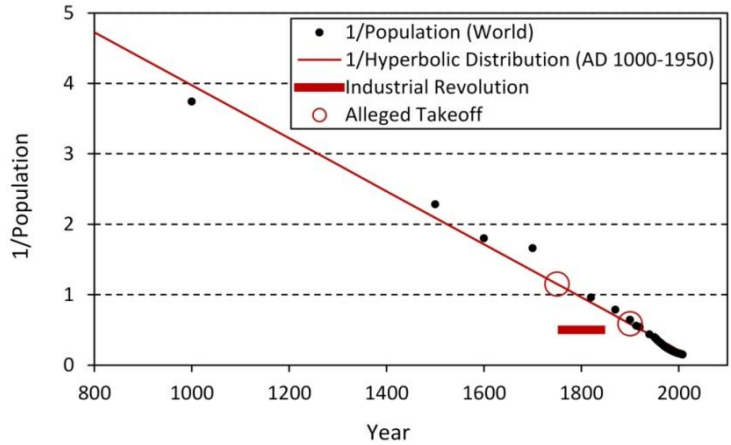

**Figure 3**. *Reciprocal values of the world population data (Maddison, 2010) identify uniquely hyperbolic distribution between AD 1000 and around 1950 because they follow a decreasing straight line. From around 1950, the growth of the population started to be diverted to a new trajectory. Industrial Revolution had no impact on changing the growth trajectory. There were also no takeoffs from stagnation to growth around the postulated times for developed and less-developed regions (Galor, 2008a, 2012a). This analysis and the absence of takeoffs in the GDP distribution (Nielsen, 2016d) show that there were no takeoffs in the income per capita (GDP/cap) distribution. Unified Growth Theory is contradicted yet again by the same data which were used but not analysed during the formulation of this theory.*

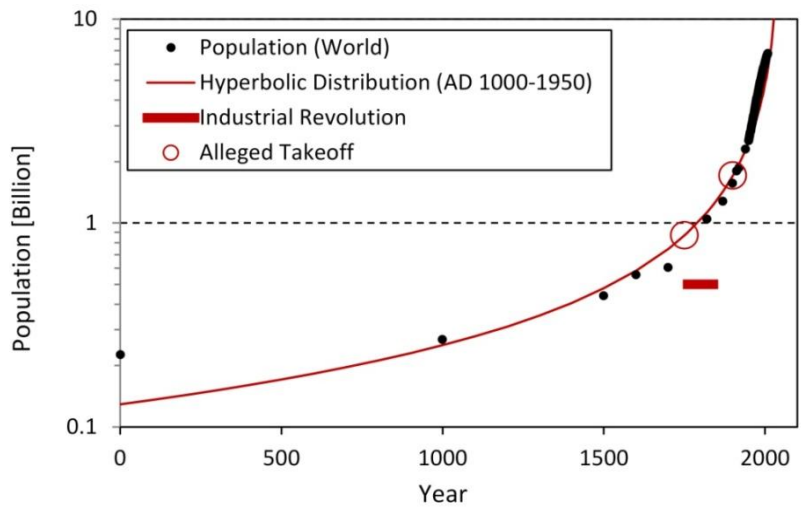

**Figure 4.** *Growth of the world population. Data of Maddison (2010) are compared with hyperbolic distribution. The point at AD 1 is 75% higher than the fitted curve because there was a maximum in the growth of the world population around that time (Nielsen, 2016c). Industrial Revolution had no impact on the growth of the population. There were no takeoffs from stagnation to growth around the postulated times (Galor, 2008a, 2012a) for developed and less-developed countries. This analysis and the absence of takeoffs in the GDP distribution (Nielsen, 2016d) show that there were no takeoffs in the income per capita (GDP/cap) distribution. Unified Growth Theory is contradicted yet again by the same data which were used but not analysed during the formulation of this theory.*

**Western Europe**

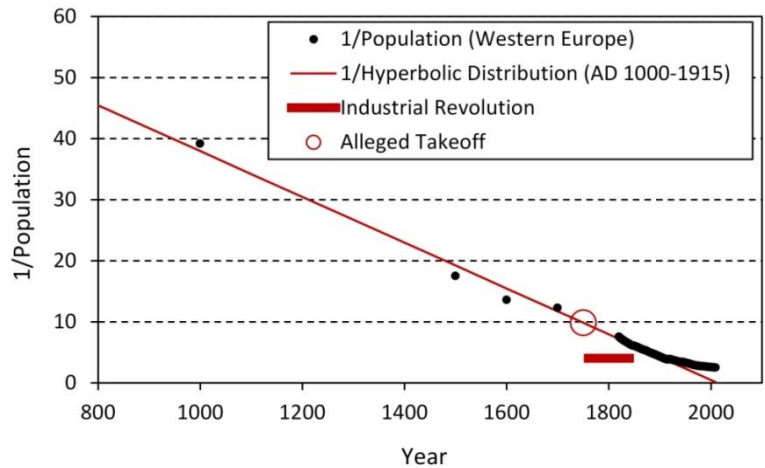

**Figure 5.** *Reciprocal values of population data for Western Europe (Maddison, 2010) identify uniquely hyperbolic distribution between AD 1000 and around 1915 because they follow a decreasing straight line. From around 1915, the growth of the population started to be diverted to a slower trajectory. Industrial Revolution had no impact on changing the growth trajectory in the region where its influence should have been most pronounced. There was also no takeoff from stagnation to growth around the postulated time (Galor, 2008a, 2012a). This analysis and the absence of the takeoff in the GDP distribution (Nielsen, 2016d) show that there was no takeoff in the income per capita (GDP/cap) distribution. Unified Growth Theory is contradicted yet again by the same data which were used but not analysed during the formulation of this theory.*

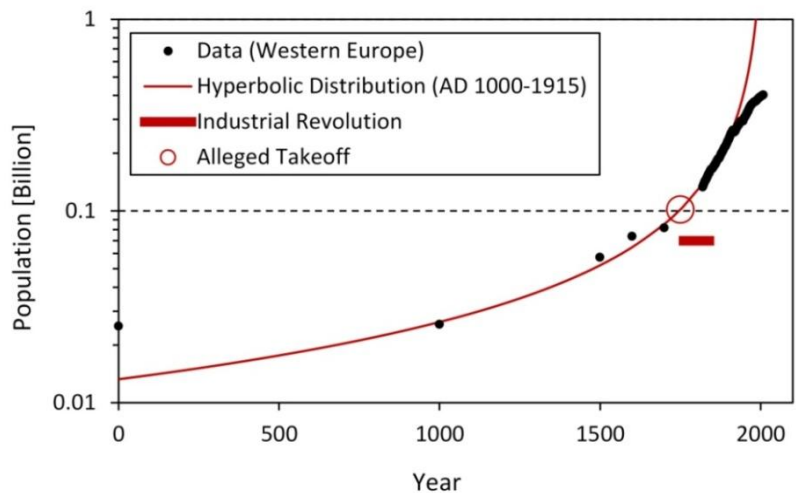

**Figure 6.** *Growth of human population in Western Europe. Data of Maddison (2010) are compared with hyperbolic distribution. The point at AD 1 is 89% higher than the fitted curve. This discrepancy might be reflecting the maximum in the growth of the world population (Nielsen, 2016c). Industrial Revolution had no impact on the growth of the population in Western Europe where the effects of this revolution should have been most prominent. There was no takeoff from stagnation to growth around the postulated time (Galor, 2008a, 2012a). This analysis and the absence of the takeoff in the GDP distribution (Nielsen, 2016d) show that there was no takeoff in the income per capita (GDP/cap) distribution. Unified Growth Theory is contradicted yet again by the same data which were used but not analysed during the formulation of this theory.*

**Eastern Europe**

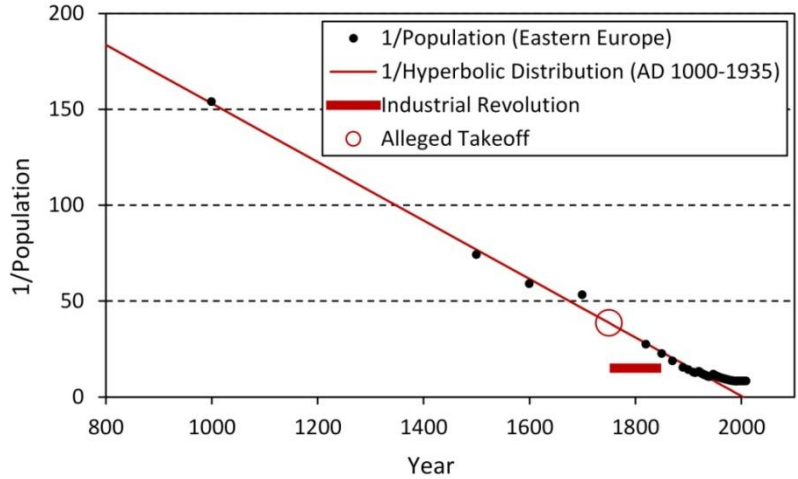

**Figure 7.** *Reciprocal values of population data for Eastern Europe (Maddison, 2010) identify uniquely hyperbolic distribution between AD 1000 and around 1935 because they follow a decreasing straight line. From around 1935, hyperbolic growth started to be diverted to a slower trajectory. Industrial Revolution had no impact on changing the growth trajectory in Eastern Europe. There was also no takeoff from stagnation to growth around the postulated time (Galor, 2008a, 2012a). This analysis and the absence of the takeoff in the GDP distribution (Nielsen, 2016d) show that there was no takeoff in the income per capita (GDP/cap) distribution. Unified Growth Theory is contradicted yet again by the same data which were used but not analysed during the formulation of this theory.*

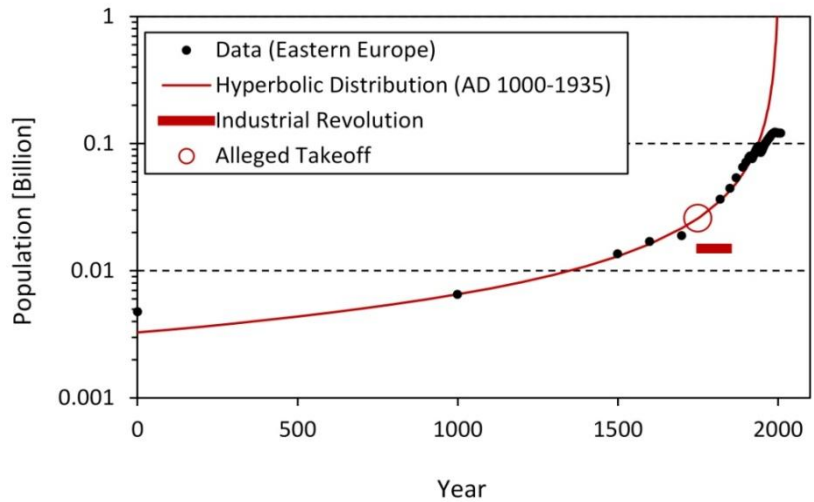

**Figure 8.** *Growth of human population in Eastern Europe. Data of Maddison (2010) are compared with hyperbolic distribution. The point at AD 1 is 45% higher than the fitted curve. This discrepancy might be reflecting the maximum in the growth of the world population (Nielsen, 2016c) around that time. Industrial Revolution had no impact on the growth of population in Eastern Europe. There was no takeoff from stagnation to growth around the postulated time (Galor, 2008a, 2012a). This analysis and the absence of the takeoff in the GDP distribution (Nielsen, 2016d) show that there was no takeoff in the income per capita (GDP/cap) distribution. Unified Growth Theory is contradicted yet again by the same data which were used but not analysed during the formulation of this theory.*

**Former USSR**

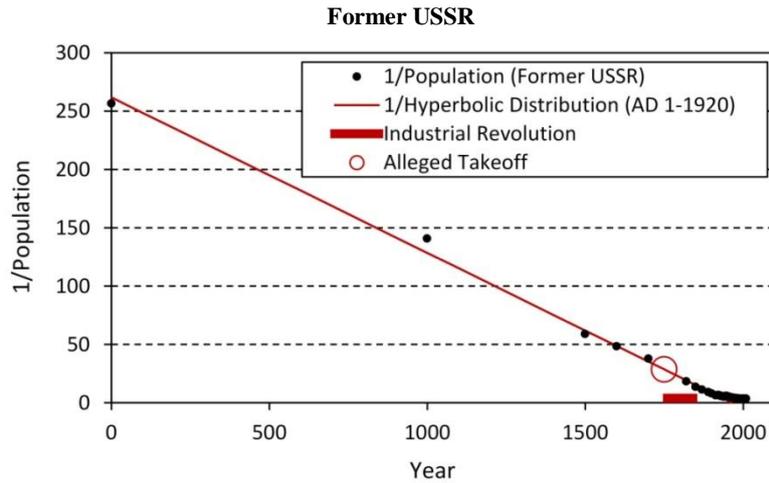

**Figure 9.** *Reciprocal values of population data for the former USSR (Maddison, 2010) identify uniquely hyperbolic distribution between AD 1 and 1920 because they follow closely the decreasing straight line. From around 1920 the growth started to be diverted to a slower trajectory. Industrial Revolution had no impact on changing the growth trajectory. There was also no takeoff from stagnation to growth around the postulated time (Galor, 2008a, 2012a) or around any other time because there was no stagnation. There was even no transition to a faster hyperbolic growth. This analysis and the absence of the takeoff in the GDP distribution (Nielsen, 2016d) show that there was no takeoff in the income per capita (GDP/cap) distribution. Unified Growth Theory is contradicted yet again by the same data which were used but not analysed during the formulation of this theory.*

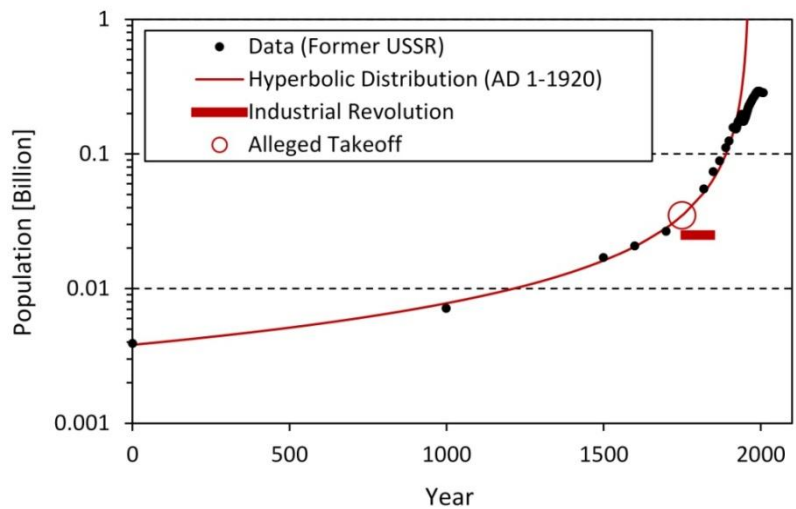

**Figure 10.** *Growth of human population in the countries of the former USSR. Data of Maddison (2010) are compared with the hyperbolic distribution. Industrial Revolution had no impact on the growth of the population. There was no takeoff from stagnation to growth around the postulated time (Galor, 2008a, 2012a) or around any other time because there was no stagnation. This analysis and the absence of the takeoff in the GDP distribution (Nielsen, 2016d) show that there was no takeoff in the income per capita (GDP/cap) distribution. Unified Growth Theory is contradicted yet again by the same data which were used but not analysed during the formulation of this theory.*

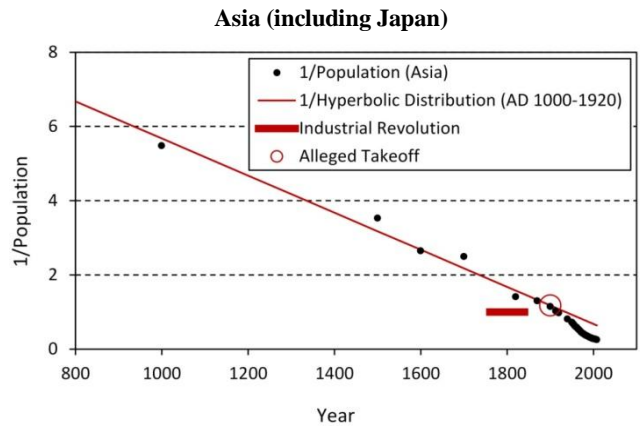

**Figure 11.** *Reciprocal values of population data for Asia (Maddison, 2010) identify uniquely hyperbolic distribution between AD 1 and 1920 because they follow closely the decreasing straight line. From around 1920, the growth started to be diverted to a temporary faster trajectory. There was no takeoff from stagnation to growth around the postulated time (Galor, 2008a, 2012a) because there was no stagnation. The temporary boosting around 1920 appears to be a part of the commonly observed transition from the historical hyperbolic growth to a slower trajectory. This analysis and the absence of the takeoff in the GDP distribution (Nielsen, 2016d) show that there was no takeoff in the income per capita (GDP/cap) distribution. Unified Growth Theory is contradicted yet again by the same data which were used but not analysed during the formulation of this theory.*

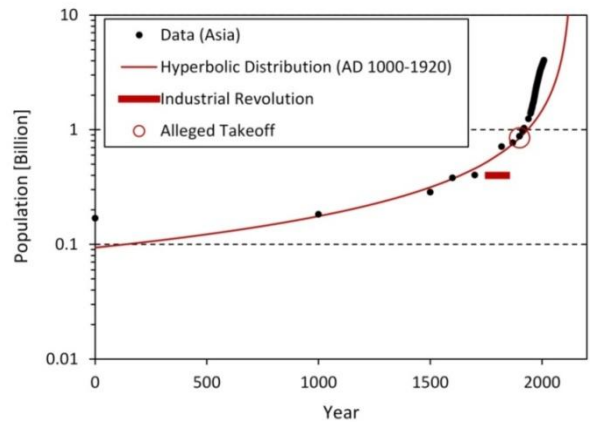

**Figure 12.** *Growth of human population in Asia. Data of Maddison (2010) are compared with the hyperbolic distribution. There was no stagnation but a hyperbolic growth between at least AD 1000 and 1920. The size of the population at AD 1 is 80% higher than the fitted hyperbolic distribution, reflecting probably the maximum in the growth of the world population around that year (Nielsen, 2016c). There was no takeoff from stagnation to growth around the postulated time (Galor, 2008a, 2012a) because there was no stagnation before the temporary boosting from around 1920. This analysis and the absence of the takeoff in the GDP distribution (Nielsen, 2016d) show that there was no takeoff in the income per capita (GDP/cap) distribution. Unified Growth Theory is contradicted yet again by the same data which were used but not analysed during the formulation of this theory.*

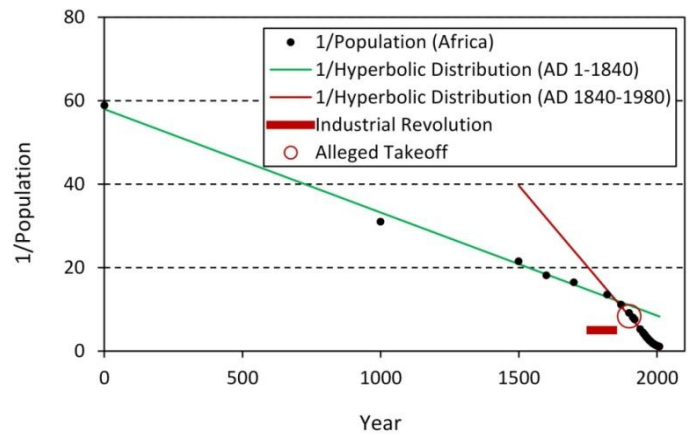

**Figure 13.** *Reciprocal values of the population data for Africa (Maddison, 2010) identify uniquely two hyperbolic distributions: AD 1-1840 and AD 1840-1980 because they follow closely the decreasing straight lines. From around 1980 the growth started to be diverted to a slower trajectory. There was no takeoff from stagnation to growth around the postulated time (Galor, 2008a, 2012a) because there was no stagnation. However there was a transition around AD 1840 from a slow to a fast hyperbolic trajectory. This analysis and the absence of the takeoff in the GDP distribution (Nielsen, 2016d) show that there was no takeoff in the income per capita (GDP/cap) distribution. Unified Growth Theory is contradicted yet again by the same data which were used but not analysed during the formulation of this theory.*

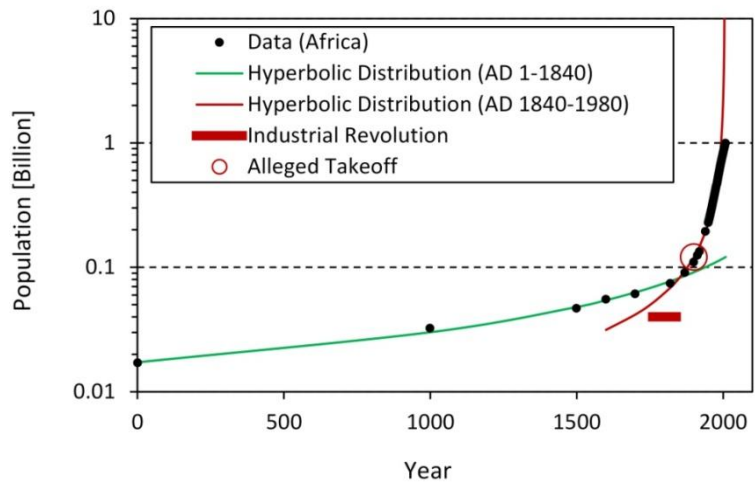

**Figure 14.** *Growth of human population in Africa. Data of Maddison (2010) are compared with two hyperbolic distributions, AD 1-1840 and AD 1840-1980. There was no stagnation but a hyperbolic growth. There was no takeoff from stagnation to growth around the postulated time (Galor, 2008a, 2012a) because there was no stagnation. The fast hyperbolic growth, continued undisturbed until 1980 when it started to be diverted to a slower trajectory. Around 1840, there was a transition from a slow to a fast hyperbolic trajectory. This analysis and the absence of the takeoff in the GDP distribution (Nielsen, 2016d) show that there was no takeoff in the income per capita (GDP/cap) distribution. Unified Growth Theory is contradicted yet again by the same data which were used but not analysed during the formulation of this theory.*

## Latin America

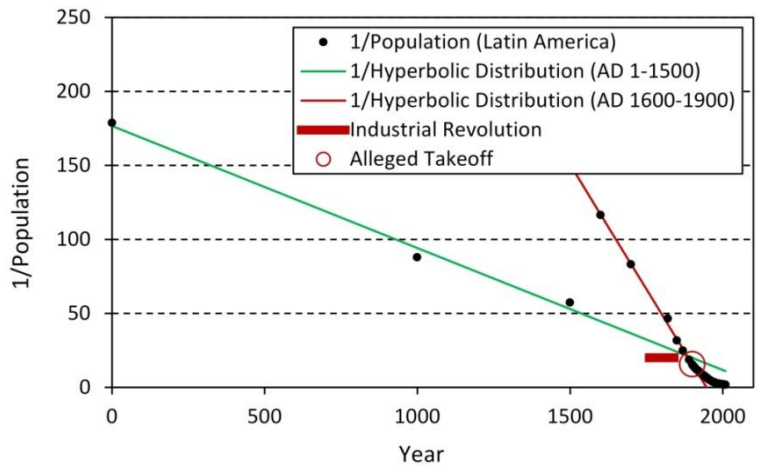

**Figure 15.** *Reciprocal values of the population data for Latin America (Maddison, 2010) identify two hyperbolic distributions: AD 1-1500 and AD 1600-1900 because they follow closely the decreasing straight lines. From around 1900 the growth started to be diverted to a slower trajectory. There was no takeoff from stagnation to growth around the postulated time (Galor, 2008a, 2012a) but there was a transition around the postulated takeoff to a slower trajectory. Data replace Galor's takeoff by a transition to a slower trajectory. This analysis and the absence of the takeoff in the GDP distribution (Nielsen, 2016d) show that there was no takeoff in the income per capita (GDP/cap) distribution. Unified Growth Theory is contradicted yet again by the same data which were used but not analysed during the formulation of this theory.*

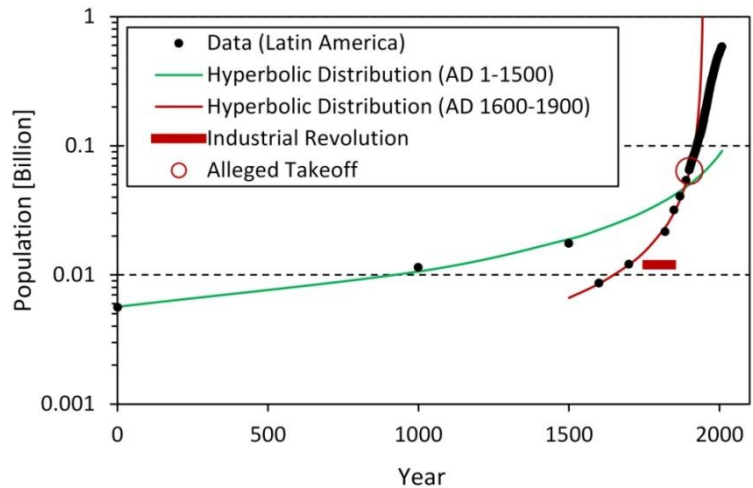

**Figure 16.** *Growth of human population in Latin America. Data of Maddison (2010) are compared with two hyperbolic distributions, AD 1-1500 and AD 1600-1900. There was no stagnation but a hyperbolic growth. There was no takeoff from stagnation to growth around the postulated time (Galor, 2008a, 2012a) because there was no stagnation. The fast hyperbolic growth continued undisturbed until 1900 when it started to be diverted to a slower trajectory. Data replace Galor's takeoff by a transition to a slower trajectory. This analysis and the absence of the takeoff in the GDP distribution (Nielsen, 2016d) show that there was no takeoff in the income per capita (GDP/cap) distribution. Unified Growth Theory is contradicted yet again by the same data which were used but not analysed during the formulation of this theory.*